\documentclass[a4paper,12pt,dvips]{article}
\usepackage{amssymb}
\usepackage{amsmath}
\usepackage{amscd}
\usepackage{epsfig}
\usepackage{multirow}
\usepackage{color}
\usepackage{subfigure}
\usepackage[a4paper=true, ps2pdf, colorlinks=true, citecolor=blue,
urlcolor=blue, linkcolor=blue, pdfstartview=Fit]{hyperref} 
\usepackage[nottoc]{tocbibind}

\usepackage[bf,footnotesize]{caption}

{\null\hfill $\Box$\par\medskip}
{\\{\bf ENDFIXME}\par\medskip}
{}

\renewcommand{\sectionmark}[1]%
{\markright{\thesection\ #1}}

\newcommand{\rf}{\mathcal{R}_5^{\mathbf{sp}}}

\def\xlf{\raisebox{+0.2em}{\color{red}\boldmath{$\chi$}}\hspace{-0.2ex}\raisebox{-0.2em}{\color{green}L}
\hspace{-1.5ex}\raisebox{+0.14em}{\color{blue}F}\hspace{2mm}}

\date{}

\begin{document}
\begin{titlepage}

\title{
  {\vspace{-0cm} \normalsize
  \hfill \parbox{40mm}{DESY 03-198\\
                       SFB/CPP-03-56\\
                       December 2003}}\\[25mm]
   Scaling test for Wilson twisted mass QCD
}

\author{{\bf \xlf Collaboration}
\\[2mm]
   Karl Jansen$^{a}$, Andrea Shindler$^{a}$,\\ 
   Carsten Urbach$^{a,b}$ and Ines Wetzorke$^{a}$\\
\\
  {\small $^a$ NIC/DESY Zeuthen} \\
  {\small Platanenallee 6, D-15738 Zeuthen, Germany} \\[5mm]
  {\small $^b$ Institut f\"{u}r Theoretische Physik, Freie Universit\"{a}t
   Berlin} \\
  {\small Arnimallee 14, D-14195 Berlin, Germany} \\
}

\maketitle

\begin{abstract}
We present a first scaling test of twisted
mass QCD with pure Wilson quarks for a twisting angle 
of $\pi/2$. We have computed the vector meson 
mass $m_{\rm V}$ and the pseudoscalar decay constant $F_{\rm PS}$
for different values of $\beta$ at fixed value of $r_0m_{\rm PS}$. 
The results obtained in the quenched approximation are compared 
with data for pure Wilson and non-perturbatively O$(a)$ improved Wilson 
computations.
We show that our results from Wilson twisted mass QCD
show clearly reduced lattice spacing
errors, consistent with O$(a)$ improvement and without the need
of any improvement terms added. These results thus provide numerical 
evidence of the prediction in ref.~\cite{Frezzotti:2003ni}.

\end{abstract}

\end{titlepage}
\setcounter{page}{2}

\section{Introduction}

Formulating QCD on a space-time lattice admits a 
substantial amount of freedom 
in discretising the continuous derivative with restrictions 
coming from
obeying principles such as gauge invariance, locality and unitarity.
Naturally, as long as universality holds, all these formulations
should provide consistent results 
in the limit when the discretisation is removed. 
The standard Wilson formulation of lattice QCD \cite{Wilson:1974sk}
is a simple realisation of such a discretised version of QCD and 
has been used for a long time in lattice simulations.
However, it has been realised that this formulation has a number 
of severe
problems: it shows large discretisation effects \cite{Jansen:1998mx} that 
are linear in the lattice
spacing $a$, it violates chiral symmetry strongly and develops 
unphysical small
eigenvalues of the corresponding lattice Wilson-Dirac operator, even at rather large values
of the quark mass.

The problem of discretisation effects can be overcome following the
Sy\-man\-zik improvement program \cite{Symanzik:1983gh,Symanzik:1983dc},
introducing the well known clover term \cite{Sheikholeslami:1985ij}
to get full, on-shell O$(a)$ improved results, if the improvement
is performed non-perturbatively \cite{Jansen:1998mx}. 
In this approach then also the chiral properties are improved,
although chiral symmetry breaking and O$(a^2)$ 
lattice spacing effects are still left in the 
theory and have to be extrapolated away.
Despite the fact that non-perturbatively improved Wilson fermions 
clearly diminish discretisation errors, they, unfortunately, show the
same -- if not worse -- 
problem of the appearance of small eigenvalues of the lattice 
Dirac operator \cite{Luscher:1997ug}.  

In order to solve the problem of small eigenvalues, it has been proposed in 
\cite{Frezzotti:2000nk}  
to use the so-called twisted mass formulation of QCD.
In this approach the mass term
in the Dirac operator is chirally twisted \cite{Frezzotti:2000nk}, see also 
\cite{Gasser:1985gg}. When using such a 
chirally twisted lattice action in combination with the clover term and 
a non-perturbatively tuned value of the improvement
coefficient, the theory is O$(a)$ improved and the 
corresponding lattice Dirac operator is safe against 
developing small eigenvalues. 

Recently, it has been realised 
\cite{Frezzotti:2003ni} that all the above properties of 
non-perturbatively improved, twisted mass QCD can also be obtained 
{\em when the clover term is completely omitted.} 
If a
special value of the twisting angle is chosen 
one
theoretically obtains O$(a)$ improved results {\it without 
adding improvement 
terms.}
At the same time,
the lattice Dirac operator
is still protected against the appearance of 
small eigenvalues by construction.
This curious observation receives a special importance for simulations
with dynamical quarks: 
In 
\cite{Aoki:2001xq} it was found that dynamical fermion simulations
with non-perturbatively improved Wilson fermions show signs of
first order phase transitions which render simulations very difficult
and induce large cut-off effects.

Although by varying the form of the gauge actions \cite{Iwasaki:1985we} 
the first
order phase transition seems to vanish, it is unclear, whether 
eventually such phenomena will reappear. 
The problem of the presence of the first order phase transition
may be related to the fact that the clover term in the fermion 
action generates an adjoint gauge action.                             
Therefore, if O$(a)$ improvement can be achieved without the clover
term, as anticipated in \cite{Frezzotti:2003ni}, the problems 
connected to such phase transitions should be completely 
eliminated.  
As a consequence,
the potential of twisted mass fermions in general 
may be very large. Since the small eigenvalues are regulated 
by the twisted mass parameter, simulations at much lower 
quark masses than used today could be performed 
with promising, but hard to estimate, advantages to explore the chiral limit
of lattice QCD. 

In this paper we provide a first test of the conjecture 
of O$(a)$ improvement of 
Wilson twisted mass QCD in quenched numerical simulations.
For this purpose we performed a scaling test of the vector meson mass $m_{\rm V}$ and
the pseudoscalar decay constant $F_{\rm PS}$ at a fixed value of the physical 
pseudoscalar mass $m_{\rm PS}$. We compare the results with those 
that have been obtained for standard 
Wilson fermions, see \cite{Gockeler:1998fn} and references therein, 
and non-perturbatively improved clover fermions 
\cite{Garden:1999fg}.


\setcounter{equation}{0}
\section{Twisted mass QCD with Wilson quarks}

In twisted mass QCD (tmQCD) as formulated in \cite{Frezzotti:2000nk}  
the twisted mass action in the continuum reads as follows:
\begin{equation}
  \label{eq:1}
  S_F[\psi,\bar \psi] = \int d^4x \bar \psi (D_\mu \gamma_\mu +m_0 + i\mu_q\gamma_5\tau^3)\psi \ ,
\end{equation}
where $D_\mu$ denotes the usual covariant derivative, $m_0$ is the
standard bare quark mass, $\tau^3$ is the third Pauli matrix acting in
flavour space and $\mu_q$ is the twisted mass parameter, also referred to as
the {\em twisted mass}. 

An axial transformation, 
\begin{equation}
  \label{eq:axialtrans}
  \psi' = \exp(i\omega\gamma_5\tau^3/2)\psi,\qquad\bar\psi'=\bar\psi\exp(i\omega\gamma_5\tau^3/2)\ ,
\end{equation}
with a real rotation angle $\omega$ leaves the 
form of the action invariant and merely changes 
the mass parameters into $m_0'$ and $\mu_q'$,
\begin{eqnarray}
m_0' & = & m_0 \cos(\omega) + \mu_q\sin(\omega) \nonumber \\
\mu_q' & = &  -m_0 \sin(\omega) + \mu_q\cos(\omega)\; .
\label{primedmasses}
\end{eqnarray}
The standard action ($\mu_q'=0$) is obtained by
setting $\tan\omega=\mu_q/m_0$. 
Note that $\tau^3$ is traceless and therefore the
transformation (\ref{eq:axialtrans}) does not couple to the 
fermion determinant
anomaly. 

In order to have an O$(a)$ improved twisted mass 
lattice action, it appears to be natural to discretise the Dirac operator
adding appropriate improvement terms 
to the standard 
Wilson-Dirac operator. 
Indeed, it has been shown in
\cite{DellaMorte:2000yp, DellaMorte:2001ys, Frezzotti:2001ea, Frezzotti:2001du} 
that by adding the usual clover term, 
full on-shell O$(a)$ improvement can be obtained. 

In \cite{Frezzotti:2003ni} it has been realised later that using 
simply the standard 
lattice Dirac operator
$D_W$ one can obtain O$(a)$ improved physical observables without 
adding improvement terms. 
More precisely, it is possible to obtain O$(a)$ improved lattice
results by employing only the standard massless 
Wilson-Dirac operator, 
\begin{equation}
D_W = \frac{1}{2} \{ \gamma_\mu (\nabla^*_\mu + \nabla_\mu) 
     - a r\nabla^*_\mu \nabla_\mu\}
\end{equation}
under the condition that one averages over physical observables that
are obtained from simulations at positive and negative values of the 
Wilson parameter $r$. 
A less general, but similar suggestion was made by the authors 
of refs. \cite{{Jacobs:1983ph},{Aoki:1984qi}}.
Instead of using positive and negative values of $r$, quark masses with 
different signs may be used: the bare quark mass $m_0$ can be written as 
\begin{equation}
  \label{eq:5b}
   m_0 = m_c(r) + m_q\ , \qquad\textrm{with}\qquad m_c(-r) = -m_c(r)\ ,
\end{equation}
where $m_c(r)$ is the critical quark mass, and $m_q$ is the subtracted bare
quark mass. Averaging physical observables obtained from simulations
at positive and negative subtracted
bare quark masses, again O$(a)$-improvement is obtained.

Let us shortly sketch the arguments leading to this surprising result.
One first has to observe that with
\begin{equation}
  \label{eq:6}
  \mathcal{R}_5 :
  \begin{cases}
    \psi\ & \to\ \psi'=\gamma_5\psi\\
    \bar\psi\ & \to\ \bar\psi'=-\bar\psi\gamma_5
  \end{cases}
\end{equation}
the following combined transformation
\begin{equation}
  \label{eq:7}
  \mathcal{R}_5^{\mathbf{sp}}\equiv \mathcal{R}_5\times[r\to-r]\times[m_q\to-m_q]
\end{equation}
is a so called {\it spurionic} symmetry of the ordinary Wilson
action. Another symmetry of the Wilson (and Wilson tmQCD) action is 
$\mathcal{R}_5 \times \mathcal{D}_d$. In the continuum, 
the transformation $\mathcal{D}_d$ has the effect of changing the 
sign of all the space-time 
coordinates and multiplies each local term $\mathcal{L}_i$ 
in the Lagrangian density by the factor $(-1)^{d[\mathcal{L}_i]}$, where
$d[\mathcal{L}_i]$ is the naive dimension of $\mathcal{L}_i$. 
The lattice version of this transformation is more involved 
and we refer to ref.~\cite{Frezzotti:2003ni} for details.

Taking now the parity properties of multiplicatively
renormalisable operators under $\rf$ and $\mathcal{R}_5 \times \mathcal{D}_d$ 
into account, one can show -- using
the Symanzik expansion -- that one
gets O$(a)$ improvement when averaging over two simulations with 
positive and negative Wilson coefficient $r$ (Wilson average
({\it WA})).
In addition,  from
the spurionic symmetry $\rf$ of the Wilson (and Wilson tm) action
one can obtain O$(a)$ improved physical observables when averaging, at a fixed value of $r$,
over two simulations with positive and negative  
$m_q$, as defined in eq. (\ref{eq:5b}) (mass average ({\it MA})), taking
into account this time the $\mathcal{R}_5$-parity of the operators.
Studying the chiral properties of the scalar condensate with
Wilson fermions, a similar suggestion was made by the authors 
in ref.~\cite{David:1984ys}.
In the special case of choosing $\omega=\pm \pi/2$, 
such an averaging procedure 
is done automatically.
A change of the sign of $r$ is equivalent to $\omega \to \omega + \pi$.
Hence, for $\omega=\pm \pi/2$ all the quantities that are
even under $\omega \to -\omega$ are automatically improved without
any averaging procedure.

It is the aim of this paper to check this conjecture in practical 
simulations by performing a scaling test for the vector meson mass
$m_{\rm V}$ and the pseudoscalar decay constant $F_{\rm PS}$ at $\omega = \pi/2$. 
The main goal is to test whether the results for $m_{\rm V}$ and 
$F_{\rm PS}$ are consistent with the anticipated leading 
O$(a^2)$ behaviour and that the
linear $a$ dependence is indeed cancelled. In addition, it is an 
interesting and important question, what the size of the remaining 
lattice spacing 
effects arising
in O$(a^2)$ will be. 

Let us list a few properties of the composite fields in the twisted mass
formulation 
before going to the numerical results.
Due to the transformation rule (\ref{eq:axialtrans}) one
also has to transform the composite fields defined in the usual 
way, 
\begin{eqnarray}
S^0(x) & = & \bar{\psi}(x)\psi(x),\;\;\;   
P^\alpha(x)  = \bar{\psi}(x)\gamma_5\frac{\tau^\alpha}{2}\psi(x),  \nonumber \\
A_\mu^\alpha(x) & = & 
           \bar{\psi}(x)\gamma_\mu\gamma_5\frac{\tau^\alpha}{2}\psi(x), 
\;\;\; 
V_\mu^\alpha(x) = \bar{\psi}(x)\gamma_\mu\frac{\tau^\alpha}{2}\psi(x)\; . 
\label{operators}
\end{eqnarray}

As an example we give here 
the relations for the axial and vector currents in the ``physical basis''
(primed quantities) and ``the twisted basis'' (unprimed quantities),
\begin{equation}
  \label{eq:3}
  A'^{\alpha}_\mu = 
  \begin{cases}
    \cos(\omega)A_\mu^\alpha+\epsilon^{3\alpha\beta}\sin(\omega)
V_\mu^\beta & \text{($\alpha=1,2$)},\\
    A_\mu^3 & \text{($\alpha=3$)},
  \end{cases}
\end{equation}
\begin{equation}
  \label{eq:4}
  V'^{\alpha}_\mu = 
  \begin{cases}
    \cos(\omega)V_\mu^\alpha+\epsilon^{3\alpha\beta}\sin(\omega)
     A_\mu^\beta & \text{($\alpha=1,2$)},\\
    V_\mu^3 & \text{($\alpha=3$)}.
  \end{cases}
\end{equation}
Note that in eq.~(\ref{eq:3}) for $\alpha=1,2$ and $\omega=\pi/2$ 
the role of the axial and vector currents are just interchanged. 
Of particular interest is the PCVC relation,
which takes the following form in the twisted basis:
\begin{equation}
  \label{eq:5}
 \partial_\mu^* V^\alpha_\mu \ = \ -2\mu_q\epsilon^{3\alpha\beta}P^\beta\ ,  
\end{equation}
where $\partial_\mu^*$ is the usual backward derivative.
Through a vector variation of the action one obtains the point-split
vector current as defined in \cite{{Frezzotti:2000nk},{Frezzotti:2001ea}}.
This current is protected against renormalisation and using the point-split 
vector current, the PCVC relation is an exact
lattice identity.
This implies that $Z_P=Z_\mu^{-1}$, where 
$Z_\mu$ is the renormalisation constant for the twisted mass 
$\mu_q$. This will become important in the extraction of
the pseudoscalar decay constant $F_{\rm PS}$ as described below. 

\section{Numerical Tests}
\setcounter{equation}{0}

In this section, we describe our numerical results for testing the
scaling behaviour of Wilson tmQCD in the quenched approximation. We
started our investigation
by performing a comparative benchmark study of different solvers 
for obtaining the quark
propagator. We found the CGS algorithm \cite{Saad:2003a} to be superior
to the 
BiCGstab and the CG algorithms. 
We therefore used the CGS algorithm throughout this work. 
Gauge field configurations were generated by standard 
heat-bath and over-relaxation techniques.  

\subsection{Mass average}

\begin{figure}[htbp]
  \centering
  \subfigure[$m_q=+0.02725, \kappa=0.16025$]{\label{fig:evpos}\input{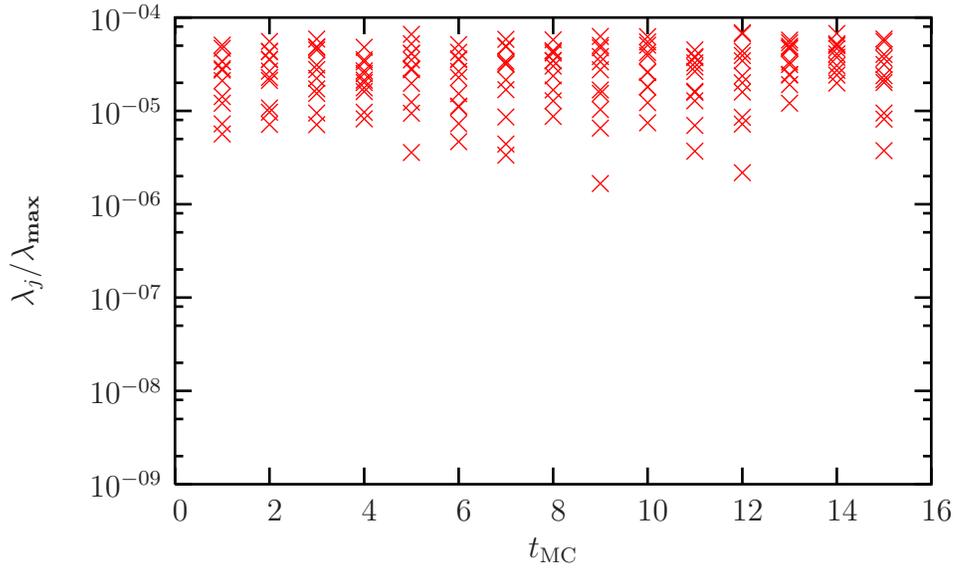}}
  \subfigure[$m_q=-0.02725, \kappa=0.163099$]{\label{fig:evneg}\input{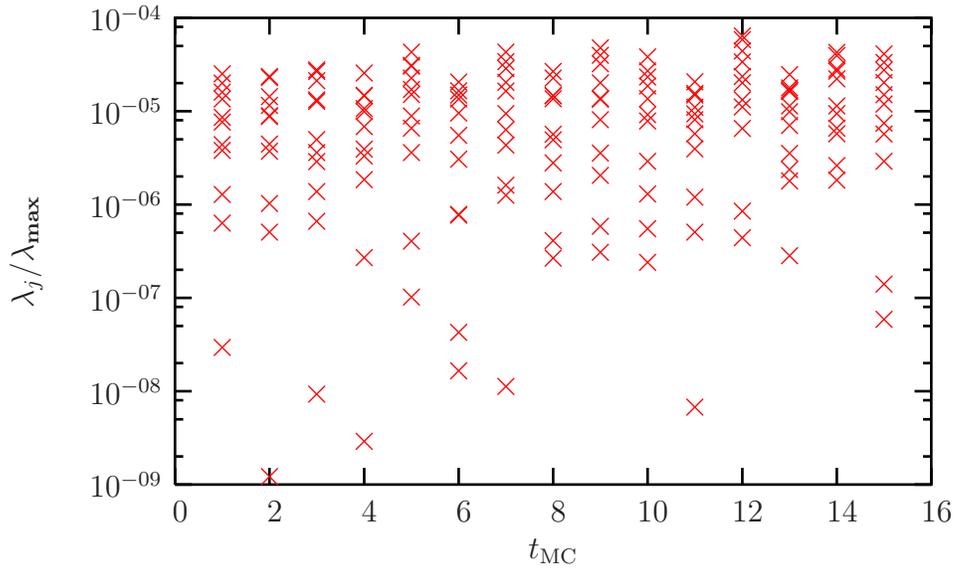}}
  \caption{ Monte Carlo time evolution of the eleven smallest
eigenvalues $\lambda$ of $(D_W+m_0)^\dagger (D_W+m_0)$, normalised by the
largest eigenvalue, at $\mu_q=0$ and 
$m_q=\pm 0.02725$ on a $12^3\times 24$
lattice ($\kappa_{c}=0.161662(17), \beta=5.85$).}
  \label{fig:ev}
\end{figure}

In order to test the predictions of ref.~\cite{Frezzotti:2003ni}, we
started with the mass average procedure. To this end, we selected a value of
$\beta=5.85$, set the Wilson parameter $r=1$ and performed simulations 
on $12^3 \times 24$ lattices at positive and negative values of
$m_q=1/2\;(1/\kappa - 1/\kappa_c)$.
While for $m_q=+0.02725$, the propagator computations went smoothly,  
for $m_q=-0.02725$ the computation of the 
quark propagator was exceedingly expensive. 
The reason for this behaviour can be traced back to the spectrum of the
Wilson tmQCD operator as can be seen in fig.~\ref{fig:ev}.

Comparing
fig.~\ref{fig:evpos} with fig.~\ref{fig:evneg}, in the case of negative 
$m_q$ one
has to deal with extremely small eigenvalues of the 
operator $(D_W+m_0)^\dagger (D_W+m_0)$. Clearly, 
these very small low-lying eigenvalues lead to a
poor convergence of the solver employed and hence to very costly
simulations. Projecting out these small modes does also not help 
in this respect since the computation of the eigenvalues is 
again costly. 

We therefore proceeded to the ``self-averaging'' case of 
choosing 
$\omega=\pi/2$, for which it has been 
shown in \cite{Frezzotti:2003ni} that 
one gets O$(a)$ improvement even without the need of
any averaging for all quantities that are even under 
$\omega \to-\omega$. 
In our practical implementation we have used the twisted basis.
Hence, the choice $\omega=\pi/2$ corresponds to set $m_q=0$ and $\mu_q\neq0$.
In this situation, 
the corresponding Wilson-Dirac
operator is protected against small eigenvalues and we do not expect 
difficulties with the simulations.

\subsection{Scaling of the vector meson mass}

In order to verify the prediction of ref.~\cite{Frezzotti:2003ni} we
computed the vector meson mass $m_{\rm V}$ and the pseudoscalar decay 
constant $F_{\rm PS}$ for the
following values of 
$\beta$: 5.85, 6.0, 6.1, 6.2. We used
periodic boundary conditions. The corresponding lattice volumes were
$14^3\times 28$, $16^3\times 32$, $20^3\times 40$ and 
$24^3\times 48$, respectively.

For our simulation in the twisted basis at $m_0=m_c$ we had to determine the
critical hopping parameter $\kappa_c$ for each value of $\beta$. 
At all the $\beta$ values of our simulations we made our own determination of 
the value of $\kappa_c$ from the intercept in $\kappa$   
at zero pion mass. The values of
$\kappa_c$ are given in table~\ref{tab:parameters}.
Note that these critical values of $\kappa$ have an 
intrinsic uncertainty of O$(a)$. 
This is, however, sufficient for obtaining fully 
O$(a)$ improved results in Wilson tmQCD \cite{Frezzotti:2003ni}. 

\begin{table}[htbp]
  \centering
  \begin{tabular}{|c|c|c||c|c|c|}
    \hline
    $\beta$ & $L$ & $T$ & $\kappa_\mathbf{c}$ & $\mu_q$ & ${\rm N}_{\rm meas}$ \\
    \hline
    $5.85$ & $14$ & $28$ & $0.161662(17)$  & $0.0376\;\;\;\,$  & 400\\
    $6.0$  & $16$ & $32$ & $0.156911(35)$  & $0.03\;\;\;\;\;\;\,\,$  & 388 \\
    $6.1$  & $20$ & $40$ & $0.154876(10)$  & $0.025854$  & 299\\
    $6.2$  & $24$ & $48$ & $0.153199(16)$  & $0.021649$  & 215\\
    \hline
  \end{tabular}
  \caption{Parameters of the simulations. Note that the values for
    $\kappa_\mathbf{c}$ are obtained from a different set of measurements.}
  \label{tab:parameters}
\end{table}

In order to fix the physical situation 
in our scaling test, 
we kept $r_0 m_{\rm PS}$ fixed for all values of $\beta$.           
For this purpose we determined the value of 
$\mu_q$ to fix $r_0 m_{\rm PS}=1.79$.
The corresponding values of $\mu_q$ for each value of $\beta$ and all our
simulation parameters are given in table~\ref{tab:parameters}.

We computed the standard 2-point correlation functions at zero momentum
for the pseudoscalar and axial operators (which in the twisted basis
at $\omega=\pi/2$ gives the correct operator to extract the
vector meson mass),
\begin{equation}
f_P^{\alpha} (t) = \sum_{\vec{x}}\langle P^\alpha(x) P^\alpha(0) \rangle \;\;\;  
\label{fP}
\end{equation}
\begin{equation}
f_A^{\alpha} (t) = \frac{1}{3} \sum_{i=1}^3\sum_{\vec{x}}\langle A_i^\alpha(x) A_i^\alpha(0) \rangle \;\;\;  
\label{fA}
\end{equation}
with $P^\alpha(x)$ and $A^\alpha(x)$ given in eqs.~(\ref{operators}) and 
$x=(\vec{x},t)$. 
In order to obtain a non-vanishing result, the flavour index 
has to be the same in these correlation functions and we will choose
$\alpha=1$ in the following. 
The pion mass could be extracted easily from the exponential 
decay of the correlation 
function $f_P^{\alpha} (t)$.   
For the vector meson mass, we performed two mass fits for the 
ground state mass 
and the first excited state. We checked the stability of the fit
by changing the value of $t_\mathrm{min}$ where the fit started. 
As a cross check, we also determined the effective ground state mass and found
consistent results. All errors were computed by a jackknife analysis. 
The numerical results at our simulation points are collected in
table~\ref{tab:results}.

\begin{table}[htbp]
  \centering
  \begin{tabular}{|c|c||c|c|c|}
    \hline
    $\beta$ & $r_0/a$ & $a m_{\rm PS}$ & $a m_{\rm V}$ & $a F_{\rm PS}$\\
    \hline
    $5.85$ & 4.067 & $0.4340(16)$ & $0.656(11)$ & $0.1147(11)$\\
    $6.0$  & 5.368 & $0.3329(21)$ & $0.488(11)$ & $0.0859(9)\;\,$ \\
    $6.1$  & 6.324 & $0.2871(17)$ & $0.427(9)\;\,$ & $0.0717(8)\;\,$ \\
    $6.2$  & 7.360 & $0.2438(16)$ & $0.363(10)$ & $0.0640(10)$ \\
    \hline
  \end{tabular}
  \caption{Results for the vector meson mass and the 
    pseudoscalar decay constant.}
  \label{tab:results}
\end{table}

\begin{figure}[htbp]
\input{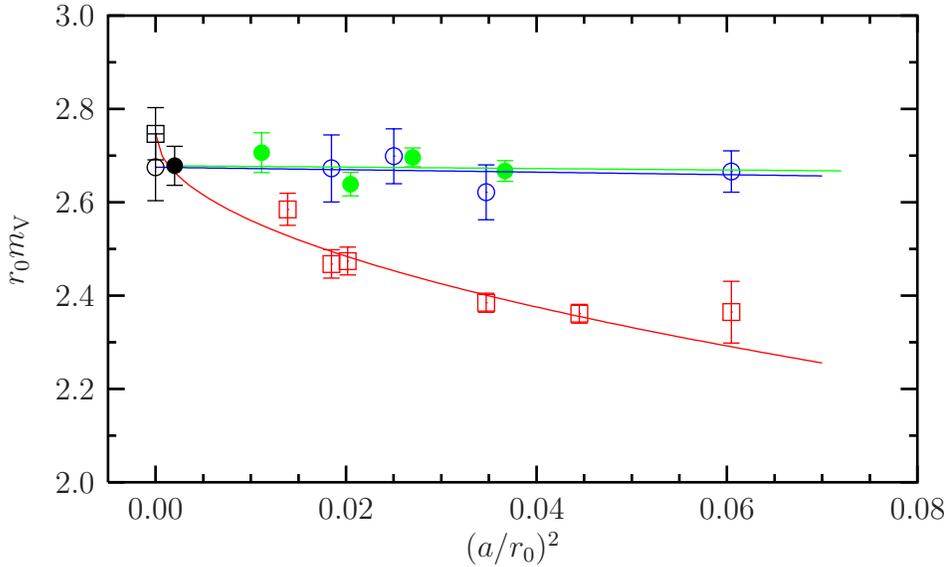}
  \caption{Scaling behaviour of the vector meson mass as
    a function of the lattice spacing squared for fixed pion mass,
    $r_0 m_{\rm PS}= 1.79$. Open circles denote Wilson tmQCD, 
    while filled circles 
    are from non-perturbatively improved Wilson fermions \protect{\cite{Garden:1999fg}}.
    The Wilson data without improvement (open squares) are collected from 
    several sources in the literature. Note that the filled circles
    are slightly displaced for better visibility.}
  \label{fig:scaling}
\end{figure}

In fig.~\ref{fig:scaling} we show our results for the vector meson mass as a 
function of $a^2$ represented by the open circles. 
In addition, we also show results from non-perturbatively
O$(a)$ improved Wilson fermions (filled circles) \cite{Garden:1999fg}. 
Finally, we added results for standard Wilson fermion simulations, 
see \cite{Gockeler:1998fn} and references therein, 
as they were available in the literature.
We remark that the published data were not always at exactly the 
same value of $r_0 m_{\rm PS}$ that we used for our Wilson tmQCD simulations.
In such cases we performed an interpolation to the desired value of
$r_0 m_{\rm PS}$. The error from this (small) interpolation is  
negligible 
for the results presented here. 

We performed a simple extrapolation of the Wilson tmQCD and the O$(a)$
improved data of 
the form $r_0 m_{\rm V} = r_0 m_{\rm V}^{\mathrm{cont}} + b \cdot (a/r_0)^2$, with 
$r_0 m_{\rm V}^{\mathrm{cont}}$ the continuum value of the vector meson mass
and $r_0 \simeq 0.5\,\textrm{fm}$. For the pure Wilson results we
replaced the quadratic term with a term 
proportional to $(a/r_0)$ in the extrapolation.
Let us remark that our data for $m_{\rm V}$ for Wilson tmQCD has about a factor
of four less statistics than the data from O$(a)$-improved Wilson fermions, 
which is reflected in the larger error bars.
Nevertheless, it is evident that the Wilson tmQCD results show a very 
similar scaling behaviour as the O$(a)$-improved Wilson fermions. 
This becomes even clearer when we compare with the unimproved pure
Wilson data for $m_{\rm V}$ that we show as open squares in 
fig.~\ref{fig:scaling}.
Here large lattice artefacts are seen and the scaling behaviour is 
much worse than with Wilson tmQCD or O$(a)$-improved Wilson fermions.
We also note that the data for Wilson tmQCD are rather flat as a function of
$a^2$ indicating that also higher order lattice spacing effects are 
suppressed. 
Clearly, it would be desirable to test these promising results in more 
precise simulations 
using a much higher statistics.

\pagebreak
\subsection{Scaling test for  
\texorpdfstring{$F_{\rm PS}$}{the pion decay constant}}

In order to extract the pion decay constant $F_{\rm PS}$ as another 
quantity to test scaling, we start with the standard definition  
of $F_{\rm PS}$ (again fixing the flavour index $\alpha=1$),
\begin{equation}
  \label{eq:deffpi}
  \langle 0|A^1_0|PS \rangle = m_{\rm PS} F_{\rm PS} \; .
\end{equation}
In the twisted basis, at $\omega =\pi/2$, the axial 
current is related to the vector current by the transformation
eq.~(\ref{eq:axialtrans}), see eqs.(\ref{eq:3},\ref{eq:4}), 
and so we can write
\begin{equation}
\label{eq:twisted_fp}
\partial_\mu\langle 0|V^2_\mu|{\rm PS} \rangle = F_{\rm PS} m_{\rm PS}^2\ .
\end{equation}
Using the vector Ward identity in eq.~(\ref{eq:5}), we can finally relate
the divergence of the vector current to the pseudoscalar density and
obtain
\begin{equation}
\label{eq:fpideterm}
 F_{\rm PS} m_{\rm PS}^2 = 
\partial_\mu\langle 0|V^2_\mu|{\rm PS} \rangle = 2\mu_q\langle 0|P^1|PS \rangle 
\; .
\end{equation}
For asymptotic Euclidean times, the pseudoscalar correlation function
$f_P^1$ assumes the form 
\begin{equation}
f_P^1(t) = \frac{| \langle 0|P^1|{\rm PS} \rangle |^2}{2m_{\rm PS}}
\cdot \left( e^{-m_{\rm PS} t} + e^{-m_{\rm PS}(T-t)}\right)\ ,
\; a \ll t \ll T\; .
  \label{eq:pp1}
\end{equation}
Thus, by fitting the pseudoscalar correlation function for large 
time separations, we can obtain $m_{\rm PS}$ and the 
amplitude $| \langle 0|P^1|{\rm PS} \rangle |^2/{m_{\rm PS}}$ from which we
then compute the desired matrix element $| \langle 0|P^1|{\rm PS} \rangle |$. 
Hence, we have all necessary ingredients to determine $F_{\rm PS}$ 
from eq.~(\ref{eq:fpideterm}), without the need of any renormalisation factor,
since $Z_P = Z_\mu^{-1}$, as we have mentioned in the previous section.
The error estimate of 
the so computed value of $F_{\rm PS}$ is performed with a jackknife 
procedure. We note that the discussion of how to get $F_{\rm PS}$ with tmQCD 
resembles very closely the strategy one would follow in the continuum or with
overlap fermions.

\begin{figure}[tbp]
\input{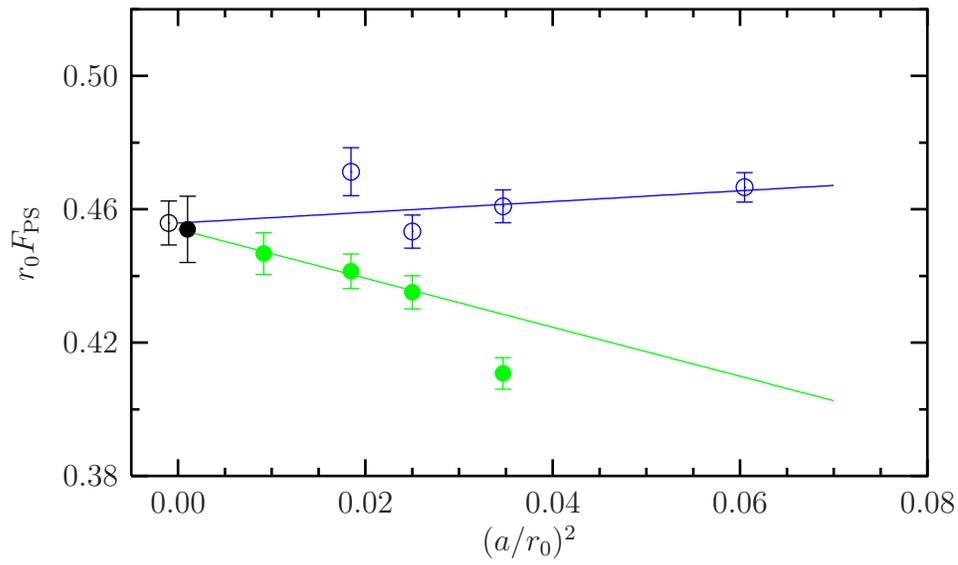}
  \caption{Scaling behaviour of the pion decay constant $F_{\rm PS}$ 
    as a function of the lattice spacing for fixed pion mass,
    $r_0m_{\rm PS}= 1.79$. Open 
    circles denote Wilson tmQCD, while filled circles 
    are from non-perturbatively improved Wilson fermions 
    \protect{\cite{Garden:1999fg}}, for which only the three data points with
    smallest values of $(a/r_0)^2$ are included in the fit. Note that the
    points corresponding to the continuum extrapolation $F_{\rm\ PS}^\mathrm{cont}$ are slightly
    displaced for better visibility.}
  \label{fig:fpi}
\end{figure}

In fig.~\ref{fig:fpi} we show our results for $r_0 F_{\rm PS}$
at a fixed value of $r_0 m_{\rm PS} = 1.79$. We also add results 
from O$(a)$-improved Wilson fermions \cite{Garden:1999fg} in the 
figure.
We performed an extrapolation of the Wilson tmQCD  and the
O$(a)$-improved data of 
the form $r_0 F_{\rm PS} = r_0 F_{\rm\ PS}^\mathrm{cont} + b' \cdot (a/r_0)^2$,
with $r_0 F_{\rm\ PS}^\mathrm{cont}$ the continuum value of the pion decay
constant in units of $r_0 \simeq 0.5\,\textrm{fm}$. While the 
O$(a)$-improved Wilson results do show cut-off effects, presumably 
due to the particular value of the improvement coefficient 
$c_A$ employed in these calculations, the results for Wilson 
tmQCD are essentially flat, indicating that also the higher order
lattice spacing effects are substantially smaller. 


\setcounter{equation}{0}
\section{Conclusion and Outlook}

In this paper we tested the striking idea of using the standard
Wilson-Dirac operator with a twisted mass and twist angle $\omega=\pi/2$ 
to obtain full O$(a)$ improvement for physical quantities that 
are even under $\omega\rightarrow -\omega$. 
The results of our -- quenched --
study for the vector meson mass $m_{\rm V}$ and the pion decay constant $F_{\rm PS}$ 
are very encouraging. It seems that in this setup the lattice spacing 
effects are substantially reduced with respect to standard Wilson fermions
and consistent with vanishing O$(a)$ 
discretisation errors. 
At the same time it seems that also lattice spacing
effects that come in higher orders in $a$ are small, a result that could not
be anticipated before. 

Thus it seems that Wilson twisted mass QCD has the potential to solve 
many problems at once:
\begin{itemize}
\item It automatically reduces lattice spacing effects,
      consistent with O$(a)$ improvement.
\item It protects against very small low-lying eigenvalues of the 
      lattice Dirac operator.
\item It can avoid unwanted phase transitions as appeared in dynamical 
      simulations with clover improved Wilson fermions and 
      Wilson gauge action.
\end{itemize}

In the light of this, dynamical simulations with twisted mass QCD 
appear to be very promising to approach the physical point at realistic 
values of quark masses. In some way, the twisted mass plays a similar role 
as an infrared cut-off as the quark mass in the staggered fermion 
or overlap approach does. 
The very good performance of staggered dynamical simulations gives hope
that also in twisted mass QCD the simulations can be accelerated as compared
to standard, improved or unimproved, Wilson fermions. Naturally, with 
Wilson twisted mass QCD the problem of unphysical taste degrees of
freedom is completely avoided, which, to us, is a clear advantage
of the tmQCD idea.

\section*{Acknowledgements}
We thank Michele Della Morte, Roberto Frezzotti and Stefan Sint for many
valuable discussions.
This work was supported in part by the European Union
Improving Human Potential Programme
under contracts 
HPRN-CT-2002-00311 (EURIDICE) and 
by the DFG through the SFB/TR9-03
(Aa\-chen-Berlin-Karlsruhe).
We are thankful to the John von Neumann-Institute for Computing 
for providing the necessary computer resources to perform this project. 


\bibliographystyle{h-physrev}
\bibliography{bibliography}
\end{document}